# Absolute Quantum Efficiency Measurements by Means of Conditioned Polarization Rotation

Giorgio Brida, Maria Chekhova, Marco Genovese, Marco Gramegna, Leonid Krivitsky, and Maria Luisa Rastello

*Abstract*—We propose a new scheme for measuring the quantum efficiency of photon counting detectors by using correlated photons. The measurement technique is based on a 90° rotation of the polarization of one photon member of a correlated pair produced by parametric down-conversion, conditioned on the detection of the other correlated photon after polarization selection. We present experimental results obtained with this scheme.

## I. Introduction

IN RECENT years, single photon detectors have found various important scientific and technological applications (e.g., quantum cryptography and laser ranging) which demand a precise determination of the quantum efficiency of these detectors. Classical calibration schemes are based on the use of a strongly attenuated source whose (unattenuated) intensity has been measured by means of a calibrated radiometer. The uncertainty of this method of measurement is mainly limited by the uncertainty in the calibration of the high insertion loss required for reaching single photon levels. A second method that allows intrinsically absolute measurements is offered by the use of correlated photons produced by type-I parametric down conversion [1]. In the process of optical parametric down-conversion, photons from a pump laser beam "decay", within a nonlinear crystal, into pairs of photons under the constraints of energy and momentum conservation. These photons are emitted simultaneously in pairs strongly correlated in direction, wavelength, and polarization. The observation of a photon on a certain direction (signal) implies the presence of another on the conjugated direction (idler), if this last is not observed this results from the non ideal quantum efficiency of the signal detector, which can be measured in this way [2], as the ratio between coincidence counts, from idler and signal detectors, and the idler counts. The systematics of this technique have been explored in order to approach high-accuracy measurements [3], [4]. Considering the high interest in the precise characterization of detectors, in this paper we propose an alternative method for the absolute quantum efficiency measurement of single photon detectors.

This work was supported by INTAS YS under Grant 03-05-1977.
G. Brida, M. Genovese, M. Gramegna, and M. L. Rastello are with the Istituto Elettrotecnico Nazionale "Galileo Ferraris," 10135 Torino, Italy.
M. Chekhova and L. Krivitsky are with the Physics Department, M.V. Lomonosov State University, 119992 Moscow, Russia.

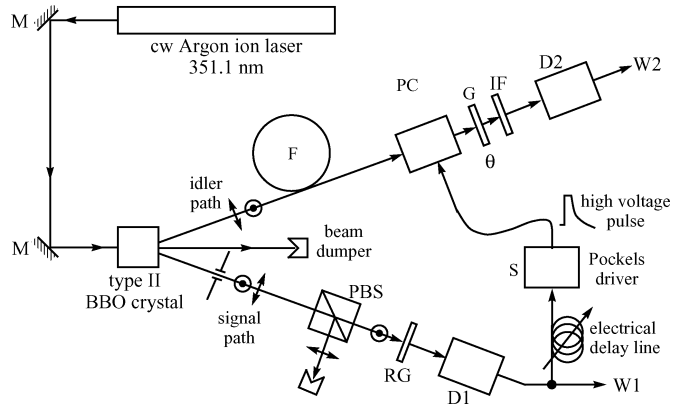

Fig. 1. Experimental set-up. The angle $\theta = 0°$ corresponds to horizontal transmission.

## II. Measurement Technique

The new proposed calibration method is based on type-II parametric down-conversion [5]. The correlated photon pair produced in type-II phase matching, in contrast to type-I phase matching, is orthogonally polarized for negative uniaxial crystals. When we have a horizontally polarized photon along the signal path the other photon of the pair, along the idler path, is vertically polarized and vice versa. One photon of the pair is then directed, after a polarization selection, to the detector D1 under calibration (Fig. 1). When this photon (signal) is detected, we rotate by 90° the polarization of the delayed (by means of optical fiber F) second photon (idler) of the pair by means of a Pockels cell (PC). In this way, the polarization state of the second photon depends on the quantum efficiency $\eta_1$ of the detector D1, since the Pockels cell is activated only when the first photon is effectively detected. The count rate $W_2$ of the detector D2 is given by the following expression:

$$W_2(\theta) = \frac{\alpha \cdot \eta_2 \cdot W_0}{2} \cdot (1 - \eta_1 \cdot \cos(2\theta)) \quad (1)$$

where $\alpha$ is the idler optical path transmittance (fiber and Pockels cell transmittance factor), $\eta_2$ is the D2 quantum efficiency, $W_0$ the rate of photon pairs, and the angle $\theta$ is the polarizer (G) setting, the angle $\theta = 0°$ corresponding to horizontal transmission. The visibility $V$ of the count rate signal $W_2$ obtained by rotating the polarizer preceding the detector D2 is a measurement of $\eta_1$

$$V = \frac{\max(W_2) - \min(W_2)}{\max(W_2) + \min(W_2)} = \eta_1. \quad (2)$$

The main aspect of this scheme is the real-time active manipulation of the polarization state of correlated photons.

## III. EXPERIMENTAL SETUP

The experimental set-up is depicted in Fig. 1. An argon-ion laser at a wavelength of 351.1 nm was incident on a nonlinear BBO crystal (5 mm × 5 mm × 5 mm) oriented for producing type-II parametric down conversion. The residual pump beam after passing through the BBO crystal is absorbed by a beam dumper. We selected from the output broadband parametric fluorescence emission the directions of correlated, frequency degenerate, photon pairs with 702 nm wavelength. After crossing a polarizing beam splitter (PBS) and selecting vertical polarization, the first (signal) photon of the pair was detected by means of a silicon avalanche single photon detector D1 preceded by a pinhole and a red glass filter (RG) (which together constitute the detection apparatus to be calibrated). The second photon (idler) of the pair was delayed by 50 m single mode (4 $\mu$m core diameter) polarization maintaining fiber (F). Input and output fiber couplings are realized with 20X microscope objectives (0.4 NA). The idler photon was then directed to a KDP Pockels cell followed by a Glan-Thompson polarizer (G), an interference filter (IF) at 702 nm (4 nm FWHM) and a second silicon avalanche single photon detector D2. When the Pockels cell driver (S), was triggered by the detector under calibration, D1, it generated a high-voltage pulse (5.2 kV) with fast rising edge (5 ns), a 180 ns flat-top and a long fall tail of 10 $\mu$s duration. The Pockels cell was operating as a halfwave plate oriented at 45° to the vertical axis. The Pockels cell controlled in this way realizes a 90° rotation on the polarization of the second photon conditioned to a measurement of a vertically polarized photon on the conjugated arm. To ensure that the flat part of the high-voltage pulse arrives at the Pockels cell simultaneously with the signal photon delayed by the optical fiber a fine, variable, electronic delay adjustment was added. Because the power dissipated by the Pockels cell driver is proportional to the trigger count rate, in order to avoid failure the internal circuit provides count rate limitation and temperature protection. In case of a fault the driver is inhibited for about 1 second. To make proper correction for the dead time of the system the counting rate on the detector under calibration must be kept smaller than $10^4$ counts per second.

## IV. EXPERIMENTAL RESULTS

The counts measured by detector D2 for an integration time of 10 s are reported (without background subtraction) in Fig. 2. When the Pockels cell was not controlled by detector D1 the behavior versus the angle orientation of the polarizer is nearly flat (squared symbol); no preference in the state of polarization is expected for the photon pairs generated by type-II parametric fluorescence, each photon pair being generated with the same probability to have a horizontally polarized photon along the signal path and a vertically polarized photon on the idler path and vice versa, a vertically polarized photon on the signal path and a horizontally polarized photon on the idler path. The small fluctuation of this curve is likely to be related to a residual coupling misalignment of the single mode polarization maintaining fiber with respect to the horizontal and vertical axes. By performing the measurement with conditioned polarization rotation (triangle symbol) we observe a superimposed sinusoidal modulation dependent on the polarizer orientation. In this case the Pockels cell effect is to flip a horizontally polarized photon into a vertically polarized one whenever a vertical polarized photon is detected by D1.

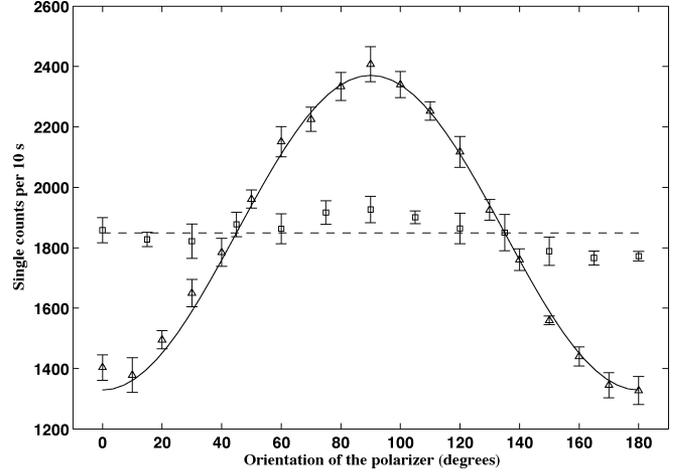

Fig. 2. Detector D2 counts in 10 s integration time, with conditioned polarization rotation (triangle) and without conditioned polarization rotation (square). Fluctuations of number of counts follows a Poisson distribution. The error bars represent the standard deviation (coverage factor, $k = 1$) of ten consecutive measurements. Lower uncertainty data have been obtained with thirty consecutive measurements. Solid line plot is the least squares fit of the experimental data for conditioned polarization rotation. Dashed line shows the mean count rate of D2 for no Pockels cell control.

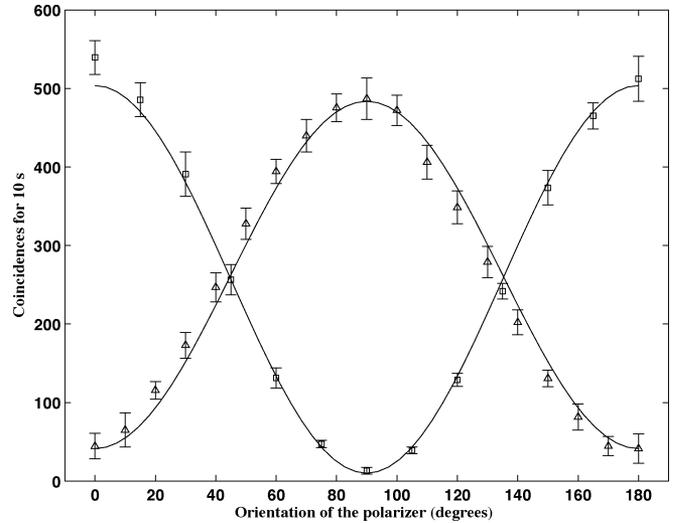

Fig. 3. Coincidences in function of the angle of the polarizer G. When the Pockels cell is not controlled by D1 the maximum is at 0° (squares). When a 90° rotation of polarization is realized by the Pockels cell conditioned to the measurement of a vertically polarized photon by D1, the maximum is shifted, as expected, at 90° (triangles). The solid line plots are least squares fits of the coincidences data with and without Pockels cell control.

In order to describe the behavior of the measurement system we report in Fig. 3 the auxiliary experimental data of coincidence events (between D1 and D2) performed with and without the conditioned polarization rotation. In this way we look selectively on the idler path only when a vertically polarized photon is detected by D1. For no Pockels cell control (square symbol) we observe a sinusoidal modulation with a maximum for horizontal polarization ($\theta = 0°$) while with the Pockels controlled by D1 detection the curve (triangle symbol) is shifted by 90° with a maximum for vertically polarized photons. From this figure we can appreciate the half-wave plate like behavior of the system from the 90° shift between the two curves.

As previously stated the visibility on the count rate signal $W_2$ obtained by rotating the polarizer G preceding the detector D2 is a measurement of the quantum efficiency of detector D1. After background estimation (by 90° rotation of the pump laser polarization) and subtraction the visibility can be evaluated from the maximum and minimum value of the recorded data (Fig. 2). The mean background rate was $640 \pm 5$ counts/s. The visibility hence evaluated has a large uncertainty (over 10%) because of the poor estimate of the count rate minimum which is comparable with its fluctuations (detector dark counts, background scattered light). Moreover, the visibility can be underestimated by an overestimation of background counts. For this reason we tried a least square adjustment (LSA) approach in order to account all the data collected in the measurement.[1] The model of adjustment considered in the evaluation task is the following:

$$W_2(\theta) = W_A \cdot \cos(2\theta) + W_B \quad (3)$$

where $W_A$ is the modulation amplitude and $W_B$ is the mean of the counting rate. The quantities $W_A$ and $W_B$ have been inferred by LSA using the measured counting rates $W_{2,j} = W_2(\theta_j)$ for different settings $\theta_j$ of the polarizer G. The angles $\theta_j$ were considered to have negligible uncertainty. The ratio between $W_A$ and $W_B$ gives the visibility. The value obtained from the experimental data by this way is $0.430 \pm 0.004$. In order to finally obtain the quantum efficiency $\eta_1$ we must correct this figure for the electrical dead times of D1 and Pockels cell driver, and for the optical losses in the polarizer cube in the signal path. The dead time correction is largely dominated by the maximum working rate of the Pockels cell driver. Its fixed dead-time is 10 $\mu$s. The corresponding correction term [4] is $0.910 \pm 0.014$ for a mean counting rate $W_1$ of $4 \cdot 10^3$ photon/s. When a further correction due to losses in the polarizer cube ($\varepsilon = 0.984 \pm 0.015$) is accounted for, we obtain the final result $\eta_1 = 0.480 \pm 0.011$ (if a small pump laser power drift could be corrected, keeping into account the change of signal counts, the result becomes $0.486 \pm 0.011$). For the sake of clarity, it must be emphasized again that the reported quantum efficiency is not the "naked" detector one, but the one corresponding to the detection apparatus including spatial and spectral filtering. In many experimental situations this is the datum necessary for understanding the performance of the set-up. If one wants to measure the "naked" detector quantum efficiency it would be necessary to introduce a corrective factor keeping into account losses in the other elements in the detection apparatus (red glass filter) and in the nonlinear crystal (corrections needed of course also for the calibration method based on correlated photons). This evaluation is beyond the purposes of this proof of principle experiment. In order to check the result obtained with the new method, we have compared it with the traditional method of single photon detector calibration by using correlated photons [2]. When we have applied the traditional correlated photons method to the same optical configuration described in the previous paragraph, our result has been $\eta_1 = 0.486 \pm 0.002$, which includes the systematic corrections [3], [4] for detector and Time to Amplitude Converter dead times. This result is close to the one obtained with the new method, $\eta_1 = 0.486 \pm 0.011$. At the moment the uncertainty is larger with the proposed technique, but a substantial reduction can be expected by a further careful study of the measurement systematics. The uncertainty in the estimate of dead-time correction and optical losses could be reasonably reduced to an insignificant level. Beside the main systematic effects that we have taken into consideration in this work some others must be accounted for, the effect of fiber coupler misalignment, the false triggering of Pockels by background counts of D1 and the dead-time correction on quantum efficiency for detector D2 [4]. Detector D2 works at very different count rates ranging in principle, as a function of the polarizer G setting, from the detector dark counts up to about $W_0$ and hence with different weighting effect. This new calibration scheme is more elaborate than the traditional calibration scheme based on correlated photons [2]. It requires a real-time feed-forward control of the polarization state of a photon, even if no postdetection electronic coincidence measurement as in [2]. Furthermore, a preliminary step is the setting of the time delay in such a way as to flip the polarization of the photon correlated with the one detected by D1. The potential attractiveness of the new scheme is the possibility to extend the method to the calibration of analog detectors.

## V. Conclusion

In conclusion, we have given a proof of principle of a new method for intrinsically absolute measurements of detector quantum efficiency based on polarization rotation of a member of a correlated photon pair conditioned to the detection of the other member after polarization selection. These first results show that the proposed method could allow an uncertainty comparable with the existing ones and, in particular, could be competitive with the one based on measurement of coincidences. Of course, a further deep investigation of all the details of this scheme and the use of a purpose built system will be necessary to reduce measurement uncertainty. The development of a new intrinsically absolute technique, which is not tied to any other standard, based in principle only upon event counting, is an independent primary standard worthy of future investigation.

## Acknowledgment

The authors wish to acknowledge fruitful discussions with Sergey Kulik (Moscow University) and Angelo Sardi (IEN).

---

[1] See "Guide to the Expression of Uncertainty in Measurement, Supplement 6. Least Squares Techniques in Metrology," prepared by members of JCGM/WG1/SC2, Nov. 24, 2003